\documentclass[a4paper,prb,reqno,twocolumn,showpacs,footinbib]{revtex4-2}
\usepackage[utf8]{inputenc}
\usepackage{amsmath}
\usepackage{amsfonts}
\usepackage{dsfont} 
\usepackage{amssymb}
\usepackage{makeidx}
\usepackage{graphicx}
\usepackage{color}

\usepackage{mathptmx} 
\usepackage{changes,enumitem,todonotes}
\usepackage{soul}
\newcommand{\bea}{\begin{eqnarray}}
\newcommand{\eea}{\end{eqnarray}}

\usepackage[mathscr]{euscript}
\usepackage{tikz}
\usepackage[normalem]{ulem}
\usepackage[utf8]{inputenc}
\usepackage{amsmath}
\usepackage{amsfonts}
\usepackage{amssymb}
\usepackage{subcaption} 
\usepackage{hyperref}
\hypersetup{colorlinks=true,linkcolor=blue,urlcolor=blue,citecolor=blue} 
\usepackage{mathtools} 
\usepackage{soul,xcolor}

\begin{document}
\setstcolor{red}

\title{Detection of long-range coherence in driven hot atomic vapors by spin noise spectroscopy}
\author{Rupak Bag, Sayari Majumder, Saptarishi Chaudhuri, and Dibyendu Roy } 
\affiliation{ Raman Research Institute, Bengaluru 560080, India}

\begin{abstract}
We study intriguing dynamical features of hot Rubidium atoms driven by two light fields. The fields resonantly drive multiple Zeeman states within two hyperfine levels, yielding a cascaded-$\Lambda$ like structure in the frequency space. A non-Hermitian Floquet tight-binding lattice with imaginary hopping between the nearest states effectively describes the coherence dynamics between Zeeman states within the ground hyperfine level.
By performing spin noise spectroscopy, we observe higher harmonic peaks in the noise spectrum arising from long-range coherence between distant ground states via multi-photon transitions. Moreover, the peak amplitudes reveal an exponential decay of the coherence with increasing separation between the states. 

\end{abstract}
\maketitle
    {\it{\textcolor{black}{Introduction}.}--}
    Optical transitions of an electron between different energy levels in atoms or molecules are constrained by selection rules. The selection rules ensure the conservation of energy and angular momentum of the electron and the (emitted or absorbed) photons \cite{Borrelli_2009} involved in the transitions. Such optical transitions are often probed using different spectroscopic techniques to gain insights into the constituent atoms and molecules of a sample near equilibrium. 
    Nevertheless, these selection rules also challenge the ability to extract  detailed information about the systems and to exert precise control over them. Driving the systems to an out-of-equilibrium condition is a simple way to overcome the hurdles posed by the selection rules \cite{DRoy_SpinNoise_PhysRevLett2014, Swar2021}. Recent years have witnessed tremendous progress in studying periodically driven systems \cite{Eckardt_RevModPhys2017} for different purposes, such as understanding thermalization and chaos \cite{Roy2022, Doug_ManybodyChaos_2023Commun}, generating and controlling topological \cite{Ozawa_SynDimension_NatRev2019,Zhou_Cavity_PhysRevLett2017, Vyas2021} and dynamical phases of matter \cite{Xu_FloquetSuperradianceLattices_PhysRevLett2022}, and quantum measurement and sensing \cite{Slim2024, Majumdar2025}. 

In this letter, we theoretically and experimentally explore intriguing dynamical features of hot vapors of $^{85}$Rb atoms under a two-tone optical driving protocol. We propose that the coherence dynamics between Zeeman states within a ground hyperfine level are governed by a non-Hermitian Floquet tight-binding lattice. Using resonant Raman fields, two neighboring Zeeman states within the ground manifold are coupled to a common excited Zeeman state in an upper hyperfine level. The effective coupling, hence generated between the two low-lying states, gives an imaginary hopping amplitude $( iJ_o)$. We capture this feature using a set of effective equations of motion describing coherence
between the ground states. The emergent energy scale is  $J_o=\Omega^2/\gamma$, where $\Omega$ is the Rabi angular frequency of the Raman fields and $\gamma~(> \Omega)$ is the population decay rate from the excited states. 

Below, we provide a theoretical model of the light-matter interaction to characterize the above-mentioned features in a driven hot vapor of $^{85}$Rb atoms. 
Based on that, we reveal that the spin noise power spectrum of the ground Zeeman states displays multiple peaks that are separated by the frequency difference between two Raman beams. These higher harmonic peaks  have their origin in the non-Hermitian  nature of the effective Hamiltonian governing the  coherence dynamics in the ground manifold.
We validate our claim by performing dispersive Faraday rotation fluctuation measurements \cite{Crooker_SpinNoise_nat2004} on the Rb atoms inside a glass cell. In the experiment, a far-detuned probe beam passes by the driven atoms without imparting significant perturbation to them and extracts information on instantaneous spin dynamics (within the ground manifold) through its polarization fluctuations \cite{MIhaila_Theory_PhysRevA2006, Zapasskii2013, Sinitsyn_SpinNoiseReview_2016}. The spin noise spectrum previously revealed many equilibrium \cite{Crooker_SpinNoise_nat2004, Oestreich2005, Muller2008, Crooker2010, Yan2012, Zapasskii_OpticalSpinNoise_PhysRevLett2013, Yang2014, roy2015cross, Lucivero_Squeezed_PhysRevA2016, Fomin_SpinAlignment_PhysRevResear2020,Gribakin_ARSpinNoise_PhysRevResearch2025, Delpy2025} and non-equilibrium \cite{pershin2013twobeam, DRoy_SpinNoise_PhysRevLett2014, Smirnov_NOneqSpinNoise_2017PhysRevB, Wiegand_QDot_PhysRevB2018, Swar2018, zhang_pulsemodulated_2020OptExp, Swar2021} features in atomic vapor and solid state systems.
For instance, when a pulsed magnetic field periodically drives an atomic ensemble, one expects 
many peaks in the spin noise spectrum due to Floquet engineering \cite{zhang_pulsemodulated_2020OptExp}. In contrast, we propose yet another optical driving protocol, where we predict and experimentally detect higher harmonic peaks in the noise spectrum, in which each harmonic characterizes the range of coherence set within the ground hyperfine manifold.

\begin{figure*}
\centering
    \subfloat{{\includegraphics[width=0.34\linewidth]{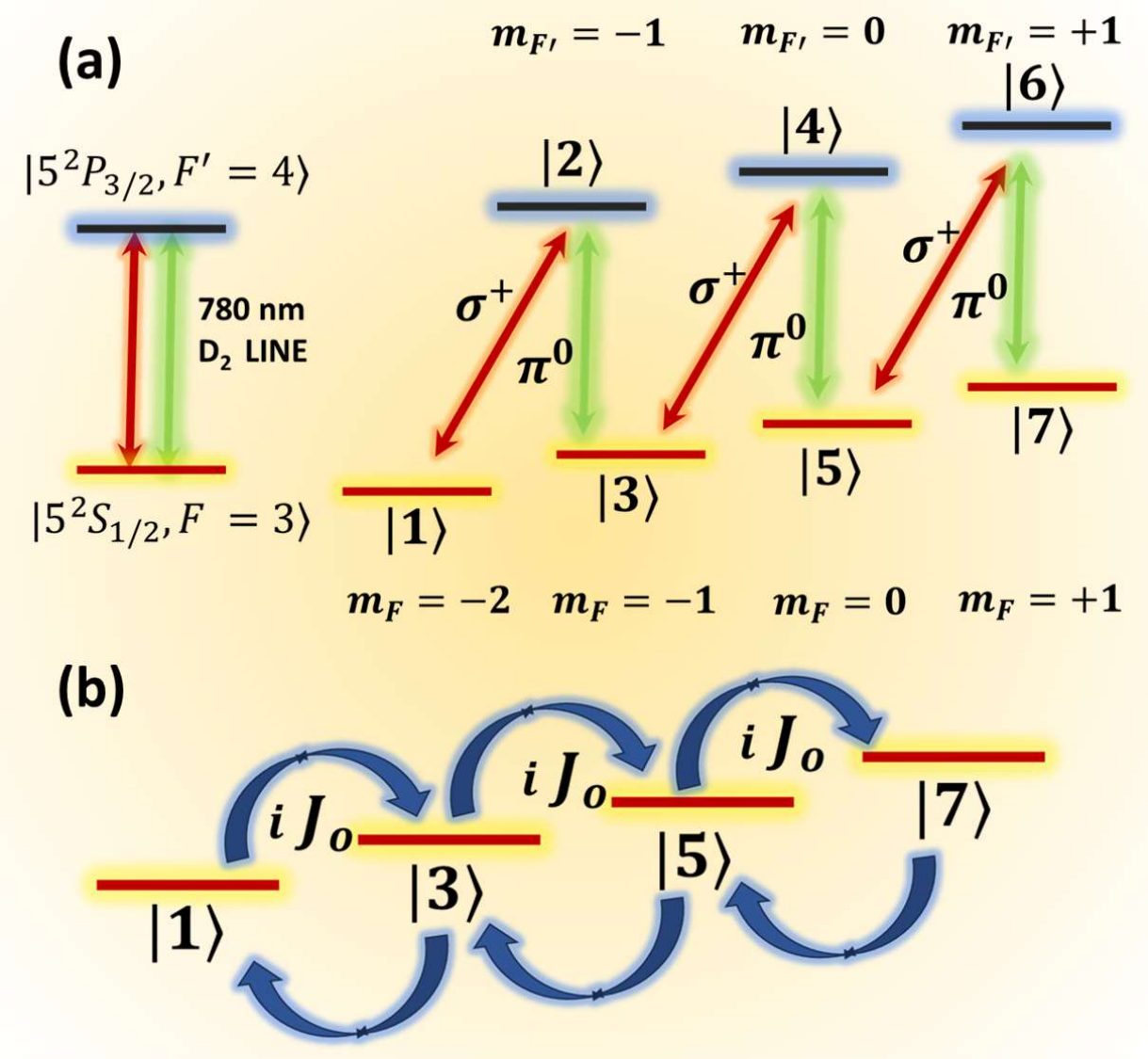}}}
    \subfloat{{\includegraphics[width=0.655\linewidth]{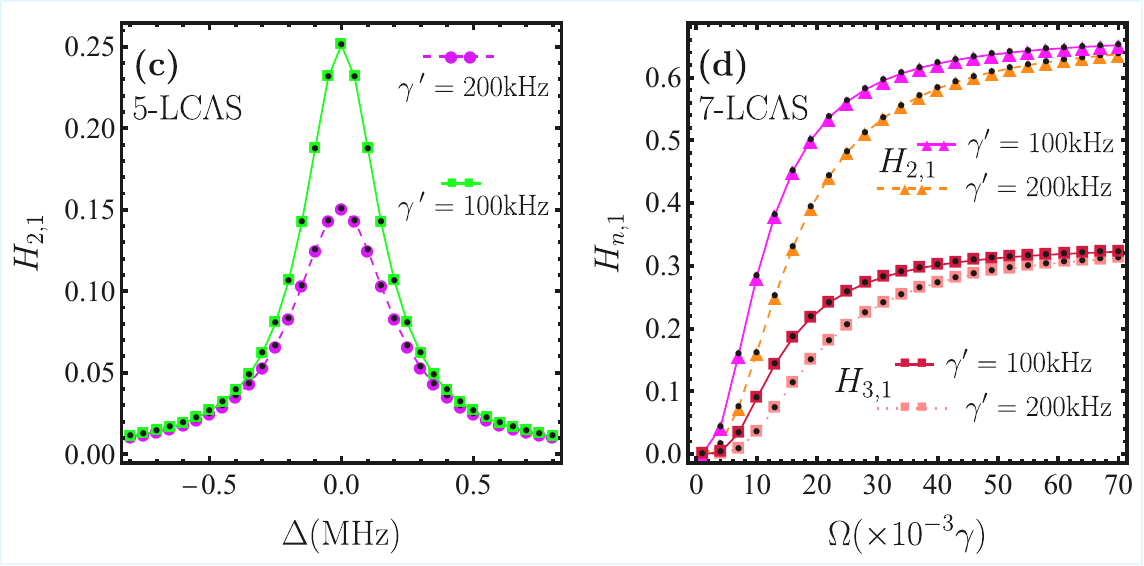} }}
\caption{(a) Energy level schematics. Hyperfine  levels, $|5\,^{2}\text{S}_{1/2}, \text{F}=3 \rangle$ and $|5\,^{2}\text{P}_{3/2}, \text{F}'=4 \rangle$ of Rb atoms coupled by two Raman fields. They drive $\pi^{0},\sigma^{+}$ transitions, which generate a C$\Lambda$ structure in the frequency space. Here, we have shown a $7$-LC$\Lambda$S. (b) The coherence between the Zeeman states in the ground hyperfine manifold is governed by an effective non-Hermitian (Floquet) tight-binding lattice with imaginary hopping $i J_o$ between the states. Such non-Hermitian dynamics generate long-range coherence between distant ground states. (c) Variation of the height-ratio $H_{2,1}$ with detuning $\Delta$ in a $5$-LC$\Lambda$S; and (d) variation of the height-ratios $H_{3,1},H_{2,1}$ in a $7$-LC$\Lambda$S with the Rabi frequency  $\Omega$. We use $\Delta_1=\Delta_2=0$ in (d) and $\gamma=1.9$ GHz in (c-d).  The solid and dashed lines are   obtained numerically using the steady state of C$\Lambda$ systems; the black dots are obtained from the effective non-Hermitian description. }
\label{Fig1}
\end{figure*}

{\it{\textcolor{black}{Theoretical  modeling}.}--} 
We consider a ground and an excited hyperfine level within the $\text{D}_2$ transition line of $^{85}$Rb atoms. For the latter part in the experiment, we choose them as $|5\,^{2}\text{S}_{1/2}, \text{F}=3 \rangle$ and $|5\,^{2}\text{P}_{3/2}, \text{F}'=4 \rangle$, respectively. An external magnetic field in the setup lifts the degeneracies of the Zeeman states in these levels. Additionally, we use two phase-coherent Raman fields to drive the atoms out of equilibrium by optically driving transitions between the ground and excited states. We are interested in a particular driving scheme, where the polarization states of the Raman fields are chosen such that two neighbouring ground Zeeman states couple to one common excited state, as shown in Fig.1(a). 
The resulting schematic looks like a cascaded-$\Lambda$ (C$\Lambda$) structure in the frequency space of the atoms.

We consider a model system with $N$ (odd) single-particle states in a C$\Lambda$ configuration to capture the essential physics. The ground and excited Zeeman states in this reduced system are represented by  $|m\rangle$, with odd and even $m$, respectively. 
The Floquet Hamiltonian describing the light-matter interaction (under rotating-wave approximation) in $N$-level C$\Lambda$ system ($N$-LC$\Lambda$S) is given by $\hat{H}(t)=\hat{H}_o+\hat{V}(t)$, where $(\hbar=1)$
\begin{align}
    &\hat{H}_o=\sum_{m= 1}^{N}{\omega}_m \hat{\sigma}_{m,m},\nonumber\\
    &\hat{V}(t) =\Omega\sum_{m= 1}^{N-1}\big(\hat{\sigma}_{m+1,m}e^{i(-1)^{m}\omega^{s}_{m}t}+\text{H.c.}\big).\label{Eq1}
\end{align}
The bare energy of a state $|m\rangle$ is $\omega_{m}$, and later we use $\omega_{m,m'}=\omega_m-\omega_{m'}$. We denote the frequencies of the two Raman fields by $\omega_{s}$ and $\omega_{s}+\delta\omega_s$; here, $\Omega$ is their  Rabi angular frequency.
In Eq.~\ref{Eq1}, $\hat{\sigma}_{m,m'}=|m\rangle \langle m'|$, and $\omega^{s}_{m}=\omega_{s}+\delta\omega_s$  $(\omega^{s}_{m}=\omega_{s})$ for odd (even)-$m$ values. 

Atoms in a vapor cell undergo collisions (with other atoms or buffer gas, or with the boundaries of the glass wall), scattering from the laser fields, and decay spontaneously from excited states. Keeping these processes in mind,  we introduce two dissipative energy scales in our modeling: (a) $\gamma$, the decay rate of population from an excited Zeeman state to a ground one, and (b) ${\gamma\,'}$ represents the intrinsic relaxation rate within the ground hyperfine manifold. We keep $\gamma \simeq 1.9$ GHz (linewidth of the excited states) and $\gamma\,' \simeq 200$ kHz. 

{\it{\textcolor{black}{Transitions in driven systems}.}--} Using time-dependent perturbation theory, we now provide a simple description of the transition rates within the ground hyperfine manifold of the model in Eq.~\ref{Eq1}. The retarded Green's function of the system including the relaxation rates $\gamma,\gamma\,'$, can be written as:
\begin{align}
 \hat{g}^{+}_o(\omega)=\sum_{m}\frac{|m\rangle \langle m|}{\omega-\omega_m+i \gamma_m}, 
\end{align} 
where $\gamma_m=\gamma$ $(\gamma_m=\gamma\,')$ for even (odd) $m$ values. Let us define the detuning of the Raman fields from an optical transition by  $\Delta_m= (-1)^{m}\omega_{m,m+1}-\omega^{s}_{m}$. Applying $\hat{g}^{+}_o(\omega)$, we derive the transition rates between different ground states for $N=7$ in \cite{SM}. When the driving is resonant, i.e., $\Delta_{m}=0$ for all $m$, the leading-order term in the transition rate from $|1\rangle \to |1+2n\rangle$ is given by:
\begin{align}
W^{(2n)}_{1+2n,1} =& (2\pi )\frac{J_o^{2n}}{\gamma\,'^{2n-2}} \delta{(\omega_{1+2n,1}-n\delta\omega_{s})} \nonumber\\
=&\mathcal{W}^{(2n)}_{1+2n,1}\delta{(\omega_{1+2n,1}-n\delta\omega_{s})},\label{Eq3}
\end{align}
where $J_o=\Omega^2/\gamma$. The  ratio of the above transition rate amplitudes varies as: 
\begin{align}
\frac{\mathcal{W}^{(2n)}_{1+2n,1} }{\mathcal{W}^{(2)}_{3,1}}=\Big(\frac{J_o}{\gamma\,'}\Big)^{2n-2}. \label{Eq4}
\end{align}

 It is obvious from Eq.~\ref{Eq3} that the higher-order transitions beyond the equilibrium selection rules $(\Delta m_F =\pm 1)$ are allowed by multi-photon transitions in the driven system. Below, we consider a more microscopic system-bath model (see \cite{SM}) to include the relaxation parameters in our description. 
  \\

{\it{\textcolor{black}{Power spectrum of coherence}.}--}  A large collisionally broadened linewidth of the excited states in atomic vapors leads to $\gamma> \Omega>\gamma\,'$, where the populations in the excited states become negligible at times $t (\gg\gamma^{-1})$.
Therefore, we only look into the spectral content of the coherence between the ground Zeeman states. We define the power spectrum between a pair of ground states  $\{|l\rangle,|l+2n\rangle\}$ using a two-time correlation function:
\begin{align}
   P_{l}^{(n)}(\omega)=&\frac{1}{2\pi} \int_{-\infty}^{\infty} d\tau~ e^{i\omega \tau} ~ \langle \hat{\sigma}^{\dagger}_{l,l+2n}(t) \hat{\sigma}_{l,l+2n} (t+\tau) \rangle,\label{Eq_PowerSpec}
\end{align}
for odd $l$ and $n=1,2,...,(N-1)/2$. 
The Zeeman states are separated by $n$ units away from each other in the ground manifold. In Eq.~\ref{Eq_PowerSpec}, the expectation $\langle ...\rangle$ is taken in an initial state of the system and baths, where we further ignore the noise coming from the baths.
At each instant $t$, we perform a unitary transformation from the laboratory frame to a rotating frame, such that:
\begin{align}
    \hat{\sigma}_{l,l+2n}(t)= e^{-in\delta\omega_s t}\hat{\sigma}^{R}_{l,l+2n}(t),
\end{align}
where the suffix $R$ stands for the rotating frame. 
The power spectrum in Eq.~\ref{Eq_PowerSpec} can be decomposed into two parts \cite{DRoy2017_nonrec, koshino2012control}: (a) coherent part appearing at an integer multiple of the difference between the driving frequencies, and (b) incoherent part generating an inelastic broadening about the coherent peaks. We neglect the contributions from the incoherent part since $\Omega/\gamma < 1$.
The expression for the power spectrum $P^{(n)}_{l}(\omega)$ becomes 
\begin{align}
  P^{(n)}_{l}(\omega) \simeq \delta(\omega-n\delta\omega_s)~ |\rho^{R}_{l+2n,l}|^2,\label{Eq7}
\end{align}
where $\rho^{R}_{l+2n,l}=\langle \hat{\sigma}^{R}_{l,l+2n}(t) \rangle$ for $t \gg \gamma\,'^{-1}$. The complete power spectrum is the sum: $P(\omega)=\sum_{n,l}P^{(n)}_{l}(\omega) $, containing $\delta$-peaks at the frequencies, $\omega=\delta\omega_s,2\delta\omega_s,..,(N-1)\delta\omega_s/2$. 
The coherence between the nearest-neighbor states in the ground manifold contributes to the first peak at $\omega=\delta\omega_s$; the next-nearest-neighbor coherence contributes to the second peak at $\omega=2\delta\omega_s$, and so on. 
Finally, gathering all contributions from different pairs of energy states in a single $\delta$-peak, we define the height ratio of the $n$-th peak to the first (fundamental) one as follows: 
\begin{align}
    H_{n,1}=\frac{\sum_{l} |\rho^{R}_{l,l+2n}|^2}{\sum_{l} |\rho^{R}_{l,l+2}|^2}~~\text{for odd}~l. \label{Eq8}
\end{align}
 For an analytical treatment in our model, we assume the population of each excited state $|m\rangle$ decays to its adjacent ground states $|m\pm 1\rangle$ at a rate of $\gamma$. Although Rb atoms decay obeying E1-selection rules ($\Delta m_F=0,\pm 1$), this qualitative description helps us to show that Eq.~\ref{Eq8} follows the same exponential behaviors as the transition rate ratios in Eq.~\ref{Eq4}. With this, we now show the emergence of non-Hermitian coherence dynamics in the ground hyperfine manifold.

{\it{\textcolor{black}{Effective non-Hermitian description}.}--} The populations in the excited states become negligible \cite{SM} at times $t \gg \gamma^{-1}$.  Therefore, we  integrate out their dynamics and obtain the effective equations of motion for the ground-state operators:
\begin{align}
    &\frac{\partial }{\partial t}\hat{\sigma}_{l,l'}(t)=i\Big(\hat{H}_{\text{eff}}(t)\hat{\sigma}_{l,l'}(t)-\hat{\sigma}_{l,l'}(t)\hat{H}^\dagger_{\text{eff}}(t)\Big)(1-\delta_{l,l'})- \tilde{\gamma}^{\,\text{eff}}_{l,l'}\nonumber\\
    &~~(1-\delta_{l,l'})\hat{\sigma}_{l,l'}(t)-\sum_{j}\bigg(\frac{\gamma\,'^{\text{eff}}_{l, j}+\gamma\,'^{\text{eff}}_{l',j}}{2}\hat{\sigma}_{l,l'}(t)-\gamma\,'^{\text{eff}}_{j,l}\delta_{l,l'}\hat{\sigma}_{j,j}(t)\bigg),\label{EffLab_Eq}
\end{align}
 for all experimentally relevant parameter regimes, where  $\Delta_m \ll \gamma$ and $\Omega<\gamma$. Here $l,l',j=1,3,...,N$, and $\delta_{l,l'}$ represents the Kronecker delta symbol. 
In Eq.~\ref{EffLab_Eq}, $\tilde{\gamma}^{\text{eff}}_{l,l'},\gamma\, '^{\,\text{eff}}_{l,l'}$ are the effective relaxation rates in the ground manifold, which are symmetric in the indices $l,l'$ (see \cite{SM}).
 
Eq.~\ref{EffLab_Eq} involves an effective non-Hermitian Hamiltonian: $\hat{H}_{\text{eff}}(t)$, which reads as
\begin{align}
    \hat{H}_{\text{eff}}(t)&=\omega_{1} \hat{\sigma}_{1,1}+\omega_{3} \hat{\sigma}_{3,3}+...+\omega_{N} \hat{\sigma}_{N,N}+\hat{V}_{\text{eff}}(t),\nonumber\\
    \hat{V}_{\text{eff}}(t)&=iJ_o\big(e^{-i\delta\omega_s t}(\hat{\sigma}_{3,1}+\hat{\sigma}_{5,3}+...+\hat{\sigma}_{N,N-2})+\text{H.c.}\big),
\end{align}
where $\hat{V}^\dagger_{\text{eff}}(t)\neq \hat{V}_{\text{eff}}(t)$. The tight-binding (Floquet) lattice Hamiltonian $\hat{H}_{\text{eff}}(t)$ with nearest-neighbor hopping amplitude $iJ_o$ between the Zeeman states [Fig.\ref{Fig1} (b)] controls the dynamics of coherence in the ground hyperfine manifold. Here, $J_o$ sets the measure of non-Hermiticity. Such dissipative coupling between two Zeeman states emerges when we integrate out the common excited state in the upper hyperfine level. The effective coupling term $\hat{V}_{\text{eff}}(t)$ captures the essence of anti-{\it{PT}} symmetry \cite{Peng_antiPTRb_2016, He_AntiPT_PhysRevA2022}, where $\{\hat{P}\hat{T},\hat{V}_{\text{eff}}(t)\}=0$. Here, parity operator $\hat{P}$ acts as:  $\hat{P} \hat{\sigma}_{l,l'}\hat{P}^{-1}=\hat{\sigma}_{N-l+1,N-l'+1}$, and time-reversal operator $\hat{T}$ changes any complex number to its conjugate.
 \begin{figure*}[t!]
    \centering
    \subfloat{{\includegraphics[width=10.3cm, height=6.3 cm]{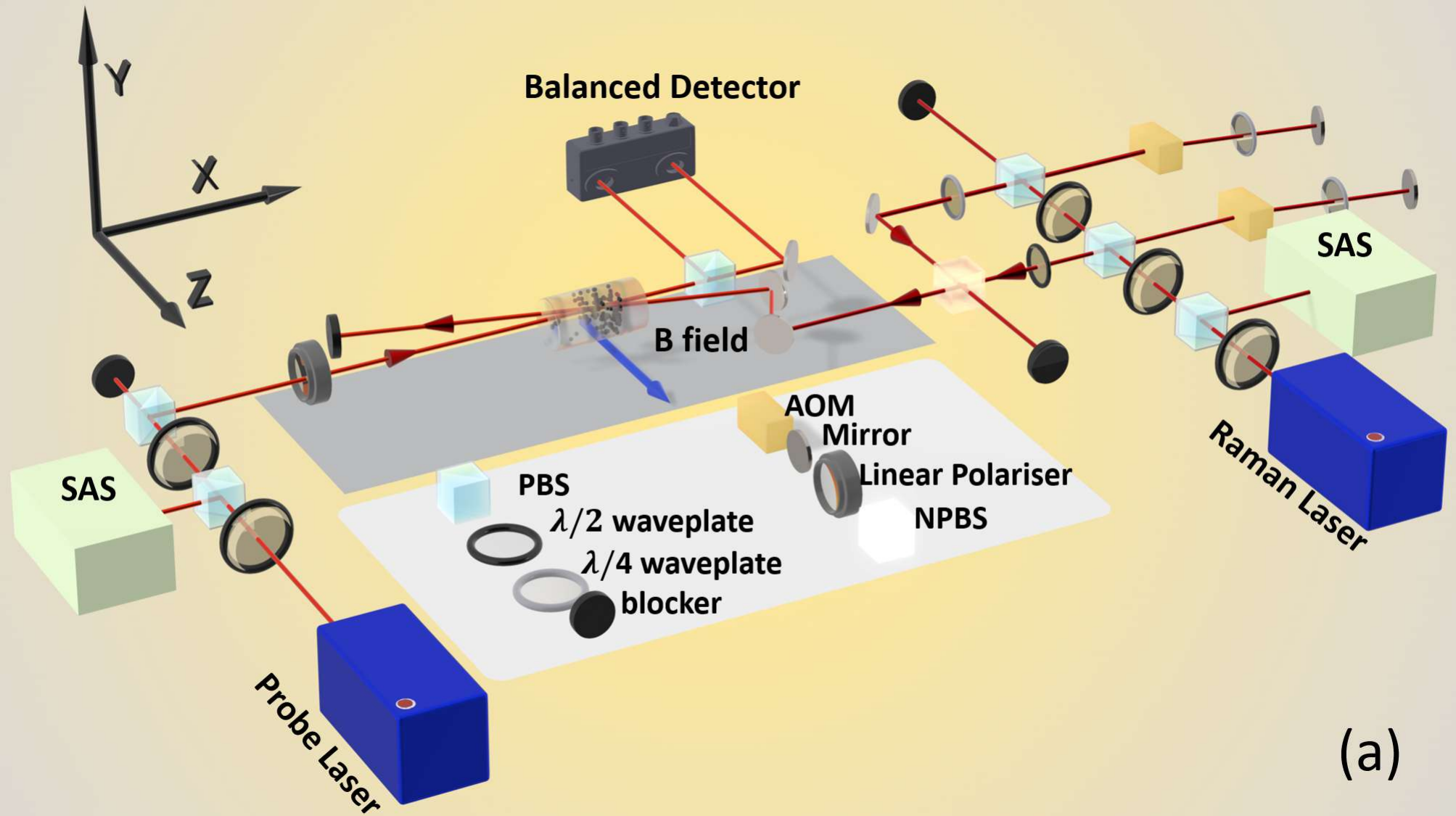}}}
    \hfill
    \subfloat{{\includegraphics[width=7.5cm,height=6.33 cm]{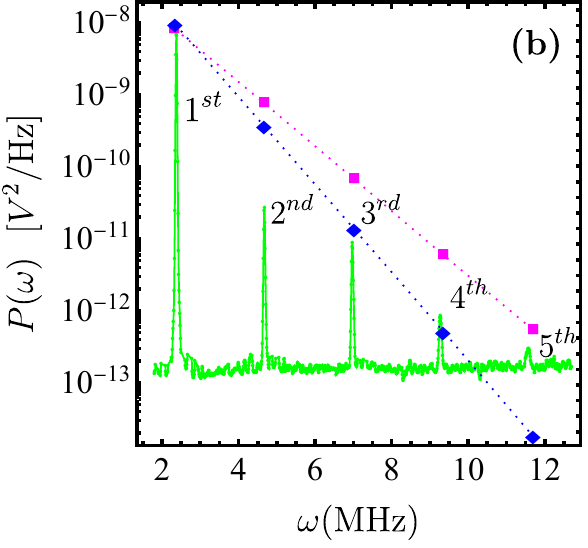} }}
\caption{ (a) The schematics for the Faraday rotation fluctuation measurements in the experiment. A linearly polarized probe beam is focused at the center of the vapor cell. A pair of Raman beams having polarizations $(\sigma^+)_x$ and $(\pi^0)_x$ is merged using a non-PBS (NPBS) and made to pass through the vapor cell in the opposite direction to the probe propagation. The frequency and intensity of the Raman fields are tuned by the AOMs. We use the saturation absorption spectroscopy (SAS) setup to determine the frequency of all laser beams. A uniform magnetic field is created along the z-axis. The probe beam falls on the polarimetric detection setup comprising a PBS and a balanced detector. (b) The experiment yields the power spectral density represented by the green-dotted line. The plot indicates five spin noise peaks in our driven system. The blue-diamond (magenta-square) points represent the corresponding theoretical comparison for the height ratios in the coherence spectrum using the full simulation of the system (using simulation of a $13$-LC$\Lambda$S). The parameters used for the simulation in the plots are: $\Delta_1=\Delta_2=0$, $\Omega=8\times 10^{-3}\gamma$, $\gamma=1.9$ GHz, and $\gamma\,'=200$ kHz. }
\label{Fig2}
\end{figure*}
Our assumption of population decay into two adjacent ground states decouples the population dynamics from the coherence terms in Eq.~\ref{EffLab_Eq}. Hence, $ \hat{\sigma}_{l,l}(t)$s remain unaffected by $\hat{H}_{\text{eff}}(t)$.  
We examine the competition between $J_o$ and others, namely $  \gamma\,'$ and $\Delta_m$, to determine the long-range coherence in the system.
We fix $\Delta_1=\Delta$ and $\Delta_2=0$, and assume the hyperfine splittings in the ground and excited Zeeman states are equal. 
To simplify our discussion, we consider three cases of $N=3,5,7$ and try to understand the trend for the higher $N$ values. In all cases, we find $\gamma\,'^{\text{eff}}_{l,l'}=J_o+\gamma\,'$ if $|l-l'|=2 $  or else $\gamma\,'^{\text{eff}}_{l,l'}=0$ for other values. 

We first consider $N=3$, where the effective Hamiltonian describes two states in the ground manifold with $\tilde{\gamma\,}^{\text{eff}}_{3,1}=J_o$. 
Following Eq.~\ref{EffLab_Eq}, the time evolution of coherence between the two states is found to depend on the sum of their populations. This feature arises from the imaginary coupling term $iJ_o$, and it plays a crucial role in sustaining a non-zero steady-state coherence between the states.
Here, the power spectrum $P(\omega)$ shows a single $\delta$-peak at the frequency $\omega=\delta\omega_s$. The amplitude of the power spectrum is proportional to $|\rho^R_{1,3}|^2=J_o^2/[(\gamma\,'+2J_o)^2+\Delta^2]$, and shows a Lorentzian fall of the height with increasing detuning $\Delta$ \cite{Swar2021}.

For $N=5$, the complicated expressions of coherence are provided in \cite{SM}. 
At lower $J_o$ values, that is, $J_o \ll \gamma\,'$, the variation of the leading-order term in the coherence with $J_o$ is as follows: $\rho^{R}_{1,3}=\rho^{R}_{3,5} \propto J_o/\gamma\,'$ and $\rho^{R}_{1,5} \propto (J_o/\gamma\,')^2$, where we take $\Delta=0$. 
The height ratio of two coherent peaks  is
\begin{align}
   H_{2,1}=\frac{2J_o^{2}}{(\gamma\,'+2J_o)^2+4\Delta^2}.
\end{align}
When  $\Delta =0$ and  $J_o\ll\gamma\,'$, we get $H_{2,1} \simeq 2 J_o^2/\gamma\,'^2$. In Fig.\ref{Fig1}(c), we display the variation of $H_{2,1}$ with $\Delta$, and find a good match between the present effective description and direct numerics of the steady-state properties of a $5$-LC$\Lambda$S.

Next, we inspect next-to-next neighbor coherence with $N=7$. We show the variations of $H_{2,1}$ and $H_{3,1}$ with increasing $\Omega$ in Fig.\ref{Fig1}(d), where  we observe the height ratios approaching saturation at large $\Omega$. Once again for  $J_o\ll\gamma\,'$ and $ \Delta=0$, we find the long-range coherence varies as follows: $\rho^{R}_{l,l+2n}\propto (J_o/\gamma\,')^n$, where the proportionality constant is different for different $n$  \cite{SM}.
 In this limit, we obtain 
\begin{align}
  H_{2,1}\simeq \frac{49 J_o^2}{ 54 \gamma\,'^2},
~~ H_{3,1}\simeq \frac{49 J_o^4}{27 \gamma\,'^4}.
\end{align}
 Earlier, we derive similar behavior of the transition rate ratios $\mathcal{W}^{(4)}_{5,1}/\mathcal{W}^{(2)}_{3,1},\mathcal{W}^{(6)}_{7,1}/\mathcal{W}^{(2)}_{3,1}$ in Eq.~\ref{Eq4}.
The above trend suggests that the peak heights of higher harmonics in the coherence power spectrum decrease almost linearly on a log-linear plot.

{\it{\textcolor{black}{Detection of coherence peaks using spin noise spectroscopy}.}--} 
We perform the experiment in a heated cylindrical glass cell containing both isotopes $^{85}$Rb and $^{87}$Rb in the presence of Ne buffer gas at $200$ Torr pressure.  To generate the probe field, we use an external cavity diode laser (ECDL) with an instantaneous linewidth of $50$ kHz. The probe frequency is blue-detuned by $+31.5$ GHz from the $F=3$ to $F'=4$ transition of the  $^{85}$Rb atoms.  The probe field is linearly polarized with a polarization purity better than $99.99\%$. It is focused through the atomic media having a waist size of $45 \mu$m, and the polarization fluctuation is detected via a polarimetric setup as shown in Fig.~\ref{Fig2} (a). The polarimetric setup comprises a polarizing beam splitter (PBS) and a balanced detector. The output of the detector is connected to a spectrum analyzer to get the power spectrum of the polarization fluctuations. We wrap the cell using ceramic wool blankets for thermal isolation, and uniformly heat it to a temperature of $(95\pm 0.05)^{\text{o}}$C. The density of Rb at this temperature is around $4\times 10^{12}$ atoms/cc.    A uniform magnetic field is applied in the direction perpendicular to the probe propagation using a pair of Helmholtz coils. To minimize the effect of the external stray magnetic field, 
we use a four-layer $\mu$-metal shield giving an isolation factor of $10^4$. 
The coils are connected to a stabilized current source with a stability better than 1 part in $10^6$. We set the magnetic field around $\sim 5$ G and observe the equilibrium spin noise spectrum with a single peak at the frequency $\omega=\omega_{3,1}=2.34$ MHz. 

 An additional ECDL generates a pair of phase-coherent Raman fields at 780 nm, which is kept resonant to the   $F=3$ to $F'=4$ transition of the $\text{D}_2$ line of $^{85}$Rb atoms \cite{Swar2021, Majumdar2025}. The exact frequencies of the lasers are measured using a wavelength meter with an absolute accuracy of $\pm 1$ MHz.   The output of this laser is split into two parts; they pass through two acousto-optic modulators (AOMs) that are kept in a double pass configuration with RF frequency tunable around $110$ MHz. A tunable frequency difference ($\delta\omega_s$) of 2-3 MHz is maintained between the AOMs to generate the two Raman fields. The polarization of the two fields are made circular $(\sigma^{+})_x$ and linear $(\pi^0)_x$ with a polarization purity $> 99.99\%$. The beams fall onto the atoms from the opposite direction to the probe propagation. We take great care so that no part of the Raman fields falls on the balanced detector in the polarimetric setup. 

In the experiment, we set $\Omega\simeq 8\times 10^{-3} \gamma$ and this yields $J_o=121.6$kHz $<\gamma\,'$.  We show the spin noise spectrum of the driven system
in Fig.~\ref{Fig2} (b), where we observe five peaks that are separated uniformly with respect to each other. The height of the first peak (fundamental) is maximum, which is located at $\omega=\delta\omega_s=\omega_{3,1}$. The height of all the higher harmonics falls rapidly towards high frequencies. The ground hyperfine level $(F=3)$ contains seven Zeeman levels; thus, we expect six peaks from the coherence spectrum. Using theoretical simulations in Fig.\ref{Fig2} (b), we find the sixth peak falls below the photon-shot noise level. First, we perform the simulation of the full system with 16 Zeeman states, considering the Hilbert space of $F=3$ and $F'=4$ levels with $E1$ decay rules [blue diamonds in Fig. \ref{Fig2} (b)]. Next, we simulate a truncated $13$-LC$\Lambda$S  with two decay channels [magenta squares in Fig. \ref{Fig2} (b)]. Both simulations show a nearly linear fall in height of higher harmonic peaks on the log-linear plot as discussed before.
The blue dotted line (which fits the diamonds) matches the experiment better than what is predicted by a $13$-LC$\Lambda$S.  
However, the linear fall indicates an underlying exponential decay of the coherence in the ground manifold, which is captured by the qualitative theory of $N$-LC$\Lambda$S. 
The saturation effects due to atomic nonlinearity occur at higher values of $J_o$, which is currently beyond the scope of our experiment.

{\it{\textcolor{black}{Conclusions}.}--}
In summary, we investigate the exciting dynamical features of a driven atomic vapor and their detection using spin noise spectroscopy. We discuss a two-tone optical driving protocol that generates long-range coherence between far-separated states in the hyperfine ground manifold. We particularly demonstrate an emergence of non-Hermitian coherence dynamics in these driven systems. We further show that spin noise spectroscopy in the system \cite{Swar2021} can probe the underlying long-range coherence manifested by higher harmonic peaks due to multi-photon transitions. Our work substantially extends the scope of detecting correlation between far-separated states in unpolarized spin systems with minimal perturbation, which directly applies to improving precision magnetometry.

{\it{\textcolor{black}{Acknowledgment}.}--}
We thank Snehal Baburao Dalvi and Maheswar Swar for their valuable early insights into this project. This work is supported by ``An initiative under the National Quantum Mission (NQM) of Department of Science and Technology (DST)", Govt. of India, and I-HUB Quantum Technology Foundation (Award no. I-HUB/SPIKE/2023-24/004).
\bibliography{references}

\onecolumngrid
\vspace{25cm}
\begin{center}
{\large \bf{Supplementary Material for ``Detection of long-range coherence in driven hot atomic vapors by spin noise spectroscopy"}} \\
\vspace{2mm}
{Rupak Bag, Sayari Majumder, Saptarishi Chaudhuri, and Dibyendu Roy} \\
\vspace{2mm}
{Raman Research Institute, Bangalore 560080, India}
\end{center}
\vspace{1cm}

\onecolumngrid
\setcounter{figure}{0}
\setcounter{equation}{0}
\setcounter{page}{1}

\section{Transitions in driven systems}
In the manuscript, we discuss a model system with $N$ (odd) single-particle states in a cascaded-$\Lambda$ configuration.
In this section of the Supplementary Material, we take $N=7$, and obtain transition rates between different ground Zeeman states using higher order perturbation theory. The interaction Hamiltonian can be explicitly written as follows: $\hat{H}(t)=\hat{H}_o+\hat{V}(t)$, where
\begin{align}
    \hat{H}_o=\sum_{m=1}^{7}\omega_m |m\rangle\langle m|,~~~~\text{and}~~~
    \hat{V}(t)=\Omega\Big( |2\rangle \langle 1|+|4\rangle \langle 3|+|6\rangle \langle 5|\Big)e^{-i\omega^{s}_{1} t}+\Omega\Big(|2\rangle \langle 3|+|4\rangle \langle 5|+|6\rangle \langle 7|\Big)e^{-i\omega^{s}_{2} t}+\text{H.c.}
\end{align}
We denote, $\omega^{s}_{1}=\omega_s+\delta\omega_s$ and $\omega^{s}_{2}=\omega_s$.
\subsection{Time-dependent perturbation theory:  $S$-matrix}
We apply time-dependent perturbation theory to obtain an analytical expression for the $S$-matrix elements [1]. Here we consider that the perturbation $\hat{V}(t)$ is turned on at time $t_o$. The time-dependent Schr{\"o}dinger  equation $(\hbar=1)$ gives
\begin{align}
    i\frac{\partial}{\partial t} |\psi(t)\rangle=\hat{H}(t)|\psi(t)\rangle,~~\Rightarrow~~\bigg(i \frac{\partial}{\partial t}-\hat{H}_o\bigg) |\psi(t)\rangle=\hat{V}(t)|\psi(t)\rangle.
\end{align}
The formal solution of the above equation with an initial condition, $|\psi(t_o)\rangle=|\phi_i\rangle$, can be written as
\begin{gather}
    |\psi^{+}(t)\rangle=|\phi(t)\rangle+\int_{t_o}^{\infty} d\tau \hat{g}^{+}_{o}(t-\tau) \hat{V}(\tau)|\psi^{+}(\tau)\rangle,\\
    \text{where}~~\hat{g}^{+}_o(t)=-i\theta(t) e^{-i\hat{H}_o t},~\text{and}~|\phi(t)\rangle=e^{-i\hat{H}_o (t-t_o)}|\phi_i\rangle=\hat{U}_o(t,t_o)|\phi_i\rangle.
\end{gather}
Using $|\psi^{+}(t)\rangle= \hat{U}(t,t_o)|\phi_i\rangle$, we obtain a perturbative series expansion for $\hat{U}(t,t_o)$ as follows:
\begin{align}
\hat{U}(t,t_o)=\hat{U}_o(t,t_o) &+\int_{t_o}^{\infty} d\tau_1 \hat{g}^{+}_{o}(t-\tau_1) \hat{V}(\tau_1)\hat{U}_o(\tau_1,t_o) + \int_{t_o}^{\infty}\int_{t_o}^{\infty} d\tau_1 d\tau_2  \hat{g}^{+}_{o}(t-\tau_1) \hat{V}(\tau_1) \hat{g}^{+}_{o}(\tau_1-\tau_2) \hat{V}(\tau_2)\hat{U}_o(\tau_2,t_o) +...
\end{align}
We define the $S$-matrix as $\hat{S}(t,t_o)=e^{i\hat{H}_ot}\hat{U}(t,t_o) e^{-i\hat{H}_ot_o}$. Thus, we get 
\begin{align}
    \hat{S}(t,t_o)
   =\hat{\mathds{I}}&-i\int_{t_o}^{t} d\tau_1 \;\hat{U}_o(0,\tau_1) \hat{V}(\tau_1)\hat{U}_o(\tau_1,0) -i\int_{t_o}^{t} d\tau_1\int_{t_o}^{\infty}  d\tau_2  \;\hat{U}_o(0,\tau_1) \hat{V}(\tau_1) \hat{g}^{+}_{o}(\tau_1-\tau_2) \hat{V}(\tau_2)\hat{U}_o(\tau_2,0) +...\nonumber\\
   =\hat{S}^{(0)}&+\hat{S}^{(1)}+\hat{S}^{(2)}+...
\end{align}
Let us define the Fourier transform of the retarded Green's function $\hat{g}^{+}_o(t)$ in the frequency domain as
\begin{align}
    \hat{g}^{+}_o(t)=\frac{1}{2\pi}\int_{-\infty}^{\infty} d\omega'\; e^{-i\omega' t}\hat{g}^{+}_o(\omega').
\end{align}
The matrix elements of $\hat{S}(t,t_o)$ in the eigenstates $|m\rangle, |n\rangle$ of $\hat{H}_o$ are represented by $\hat{S}_{m,n}(t,t_o)=\langle m| \hat{S}(t,t_o)|n\rangle$. In the following calculations, we take $t_o \to -\infty$.
\begin{align}
    {S}^{(0)}_{m,n}=&\delta_{m,n},\\
    {S}^{(1)}_{m,n}(t)=&-i \lim_{t_o\to -\infty}\int_{t_o}^{t} d\tau_1 \;e^{i\omega_{m,n}\tau_1} \langle m|\hat{V}(\tau_1)|n\rangle,\\
    {S}^{(2)}_{m,n}(t)
    =&-i \lim_{t_o\to -\infty}\int_{t_o}^{t} d\tau_1\int_{t_o}^{\infty}  d\tau_2 e^{i\omega_m \tau_1} e^{-i\omega_n \tau_2} \; \frac{1}{2\pi}\int_{-\infty}^{\infty} d\omega'\; e^{-i\omega' (\tau_1-\tau_2)}\langle m| \hat{V}(\tau_1) \hat{g}^{+}_o(\omega')\hat{V}(\tau_2)|n\rangle.\\
    {S}^{(3)}_{m,n}(t)
    =&-i \lim_{t_o\to -\infty}\int_{t_o}^{t} d\tau_1 \int_{t_o}^{\infty} \int_{t_o}^{\infty}  d\tau_2 d\tau_3  \; e^{i\omega_m \tau_1} e^{-i\omega_n \tau_3} \; \frac{1}{(2\pi)^2}\int_{-\infty}^{\infty} \int_{-\infty}^{\infty} d\omega'_1 d\omega'_2 \;e^{-i\omega'_1 (\tau_1-\tau_2)} \nonumber\\
    &~~~~~~~~~\times e^{-i\omega'_2 (\tau_2-\tau_3)}\langle m| \hat{V}(\tau_1) \hat{g}^{+}_o(\omega'_1)\hat{V}(\tau_2) \hat{g}^{+}_o(\omega'_2)\hat{V}(\tau_3)|n\rangle,\\
    {S}^{(4)}_{m,n}(t)
    =&-i \lim_{t_o\to -\infty}\int_{t_o}^{t} d\tau_1 \int_{t_o}^{\infty} \int_{t_o}^{\infty}\int_{t_o}^{\infty}  d\tau_2 d\tau_3 d\tau_4 \; e^{i\omega_m \tau_1} e^{-i\omega_n \tau_4} \; \frac{1}{(2\pi)^3}\int_{-\infty}^{\infty} \int_{-\infty}^{\infty} \int_{-\infty}^{\infty} d\omega'_1 d\omega'_2 d\omega'_3 \;e^{-i\omega'_1 (\tau_1-\tau_2)} \nonumber\\
    &~~~~~~~~~\times e^{-i\omega'_2 (\tau_2-\tau_3)}e^{-i\omega'_3 (\tau_3-\tau_4)}\langle m| \hat{V}(\tau_1) \hat{g}^{+}_o(\omega'_1)\hat{V}(\tau_2) \hat{g}^{+}_o(\omega'_2)\hat{V}(\tau_3)\hat{g}^{+}_o(\omega'_3)\hat{V}(\tau_4)|n\rangle,
\end{align}
and so on. Here, $\delta_{m,n}$ is the Kronecker delta symbol, and we use $\omega_{m,n}=\omega_{m}-\omega_{n}$. The poles of the system's retarded Green's function include the effect of the relaxation rates $\gamma,\gamma\,'$:
\begin{align}
\hat{g}^{+}_o(\omega)=\sum_{m'}\frac{|m'\rangle \langle m'|}{\omega-\omega_{m'}+i \gamma_{m'}}, 
\end{align} 
where $\gamma_{m'}=\gamma$ $(\gamma_{m'}=\gamma\,')$ for even (odd) $m'$ values. 
\subsection{Derivation of the transition rates in the ground hyperfine manifold}
\textbf{Transition rate from $|1\rangle$ to $|3\rangle$:} We first calculate $S$-matrix element between $|1\rangle$ and $|3\rangle$. We find $S^{(1)}_{3,1}=0$. The leading-order nonzero contribution comes from $\hat{S}^{(2)}$:
\begin{align}
    {S}^{(2)}_{3,1}(t)
    =&-i \lim_{t_o\to -\infty}\int_{t_o}^{t} d\tau_1\int_{t_o}^{\infty}  d\tau_2 e^{i\omega_3 \tau_1} e^{-i\omega_1 \tau_2} \; \frac{1}{2\pi}\int_{-\infty}^{\infty} d\omega'\; e^{-i\omega' (\tau_1-\tau_2)}\langle 3| \hat{V}(\tau_1) \hat{g}^{+}_o(\omega')\hat{V}(\tau_2)|1\rangle\nonumber\\
    =&-i\lim_{t_o\to -\infty}\int_{t_o}^{t} d\tau_1 \int_{t_o}^{\infty}  d\tau_2 e^{i\omega_3 \tau_1} e^{-i\omega_1 \tau_2} \; \frac{1}{2\pi}\int_{-\infty}^{\infty} d\omega'\; e^{-i\omega' (\tau_1-\tau_2)}\frac{\langle 3| \hat{V}(\tau_1)|2\rangle \langle 2|\hat{V}(\tau_2)|1\rangle}{\omega'-\omega_2+i\gamma}\nonumber\\
    =&-i \frac{\Omega^2  }{(\omega^{s}_{1}+\omega_1)-\omega_2+i\gamma}\lim_{t_o\to -\infty}\int_{t_o}^{t} d\tau_1\;e^{i(\omega_3-\delta\omega_s-\omega_1)\tau_1} \label{TRates}.
\end{align}
The prefactor $\Omega^2$  signifies a two-photon process. We can further simplify it as
\begin{align}
{S}^{(2)}_{3,1}(t \to \infty)
    = -(2\pi i)\bigg( \frac{\Omega^2}{\omega^{s}_{1}-\omega_{2,1}+i\gamma} \bigg)\delta{(\omega_{3,1}-\delta\omega_{s})}.
\end{align}
We apply Eq.~\ref{TRates} to obtain the transition rate. The probability of the transition from $|1\rangle \to |3\rangle$ is
\begin{align}
    P^{(2)}_{3,1}(t)=&|{S}^{(2)}_{3,1}(t)|^2=\bigg| \frac{\Omega^2}{\omega^{s}_{1}-\omega_{2,1}+i\gamma} \bigg|^2 \lim_{t_o \to -\infty} \int_{t_o}^{t} d\tau_1 \int_{t_o}^{t}  d\tau_2 e^{i(\omega_{3,1}-\delta\omega_{s})(\tau_1-\tau_2)}.
\end{align}
The domain of integration for $\tau_1, \tau_2$ is following: $t_o \leq \tau_1 \leq t $, and $t_o \leq \tau_2 \leq t$. We define new variables: $T=(\tau_1+\tau_2)/2$ and $\tau=\tau_1-\tau_2$. The integration domain for the new variables is: $-|t-t_o|\leq \tau \leq |t-t_o|$ and $t_o\leq T\leq t$ $(\because t_o <0, t>0)$. Using the Jacobian of transformation, we find: $d\tau_1 d\tau_2=dT d\tau$. Therefore, 
\begin{align}
    P^{(2)}_{3,1}(t)
    &=\bigg| \frac{\Omega^2}{\omega^{s}_{1}-\omega_{2,1}+i\gamma} \bigg|^2 \lim_{t_o \to -\infty} \int_{t_o}^{t} dT \int_{-|t-t_o|}^{|t-t_o|}  d\tau e^{i(\omega_{3,1}-\delta\omega_{s})\tau}=(2\pi )\bigg| \frac{\Omega^2}{\omega^{s}_{1}-\omega_{2,1}+i\gamma} \bigg|^2 \lim_{t_o \to -\infty} (t-t_o)  \delta{(\omega_{3,1}-\delta\omega_{s})}.
\end{align}
The transition rate is
\begin{align}
    W^{(2)}_{3,1}=\frac{d }{dt} P^{(2)}_{3,1}(t)=(2\pi )\bigg| \frac{\Omega^2}{\omega^{s}_{1}-\omega_{2,1}+i\gamma} \bigg|^2   \delta{(\omega_{3,1}-\delta\omega_{s})}.\label{W1}
\end{align}
\textbf{Transition rate from $|1\rangle$ to $|5\rangle$:}
Here, we follow the steps of the previous calculations. We find $S^{(1)}_{5,1}=S^{(2)}_{5,1}=S^{(3)}_{5,1}=0$. The leading-order nonzero contribution is:
\begin{align}
  & {S}^{(4)}_{5,1}(t \to \infty)
    = -(2\pi i)\Bigg[ \frac{\Omega^4}{\splitfrac{(2\omega^{s}_{1}-\omega^{s}_{2}-\omega_{4,1}+i\gamma)}{(\delta\omega_{s}-\omega_{3,1}+i\gamma\,')(\omega^{s}_{1}-\omega_{2,1}+i\gamma)}} \Bigg]\delta{(\omega_{5,1}-2\delta\omega_{s})},~~\text{and}\\
   &W^{(4)}_{5,1}=\frac{d }{dt} |S^{(4)}_{5,1}(t)|^2=(2\pi )\Bigg|\frac{\Omega^4}{\splitfrac{(2\omega^{s}_{1}-\omega^{s}_{2}-\omega_{4,1}+i\gamma)}{(\delta\omega_{s}-\omega_{3,1}+i\gamma\,')(\omega^{s}_{1}-\omega_{2,1}+i\gamma)}} \Bigg|^2\delta{(\omega_{5,1}-2\delta\omega_{s})}. \label{W2}
\end{align}
\textbf{Transition rate from $|1\rangle$ to $|7\rangle$:} We find $S^{(n)}_{7,1}=0$ for $n=1,2,..,5$. The leading-order nonzero contribution is 
\begin{align}
  & {S}^{(6)}_{7,1}(t \to \infty)
    = -(2\pi i)\Bigg[ \frac{\Omega^6 }{\splitfrac{(\omega_{1,6}-2\omega^{s}_{2}+3\omega^{s}_{1}+i\gamma)(2\delta\omega_{s}-\omega_{5,1}+i\gamma\,')}{(\omega_{1,4}-\omega^{s}_{2}+2\omega^{s}_{1}+i\gamma)(\delta\omega_{s}-\omega_{3,1}+i\gamma\,')(\omega^{s}_{1}-\omega_{2,1}+i\gamma)}} \Bigg]\delta{(\omega_{7,1}-3\delta\omega_{s})},\\
   & W^{(6)}_{5,1}=\frac{d }{dt} |S^{(6)}_{7,1}(t)|^2=(2\pi )\Bigg| \frac{\Omega^6 }{\splitfrac{(\omega_{1,6}-2\omega^{s}_{2}+3\omega^{s}_{1}+i\gamma)(2\delta\omega_{s}-\omega_{5,1}+i\gamma\,')}{(\omega_{1,4}-\omega^{s}_{2}+2\omega^{s}_{1}+i\gamma)(\delta\omega_{s}-\omega_{3,1}+i\gamma\,')(\omega^{s}_{1}-\omega_{2,1}+i\gamma)}} \Bigg|^2\delta{(\omega_{7,1}-3\delta\omega_{s})}. \label{W3}
\end{align}
For the resonant driving case, we have $\omega_{2,1}=\omega_{4,3}=\omega_{6,5}=\omega^{s}_{1}$ and $\omega_{2,3}=\omega_{4,5}=\omega_{6,7}=\omega^{s}_{2}$. This gives
\begin{align}
 &W^{(2)}_{3,1}= (2\pi )\bigg( \frac{\Omega^2}{\gamma}\bigg)^2   \delta{(\omega_{3,1}-\delta\omega_{s})}=(2\pi ) J_o^2   \delta{(\omega_{3,1}-\delta\omega_{s})}=\mathcal{W}^{(2)}_{3,1}\delta{(\omega_{3,1}-\delta\omega_{s})},\\
&W^{(4)}_{5,1}= (2\pi )\Bigg(\frac{\Omega^4/\gamma^2}{\gamma\,'} \Bigg)^2\delta{(\omega_{5,1}-2\delta\omega_{s})}=(2\pi )\Bigg(\frac{J_o^2}{\gamma\,'} \Bigg)^2\delta{(\omega_{5,1}-2\delta\omega_{s})}=\mathcal{W}^{(4)}_{5,1}\delta{(\omega_{5,1}-2\delta\omega_{s})},\\
&W^{(6)}_{7,1}= (2\pi )\Bigg(\frac{\Omega^6/\gamma^3}{\gamma\,'^{2}} \Bigg)^2\delta{(\omega_{7,1}-3\delta\omega_{s})}=(2\pi )\Bigg(\frac{J_o^3}{\gamma\,'^{2}} \Bigg)^2\delta{(\omega_{7,1}-3\delta\omega_{s})}=\mathcal{W}^{(6)}_{7,1}\delta{(\omega_{7,1}-3\delta\omega_{s})},
\end{align}
where $J_o=\Omega^2/\gamma$. We therefore get the following.
\begin{align}
 \frac{\mathcal{W}^{(4)}_{5,1}}{\mathcal{W}^{(2)}_{3,1}}=\Bigg(\frac{J_o}{\gamma\,'} \Bigg)^2,~~\text{and}~~~\frac{\mathcal{W}^{(6)}_{7,1}}{\mathcal{W}^{(2)}_{3,1}}=\Bigg(\frac{J_o}{\gamma\,'} \Bigg)^4.
\end{align}

\section{Operators in the rotating frame}
In the manuscript, we facilitate a unitary transformation using $\hat{\mathcal{U}}(t)=e^{i\hat{G}t}$ with $\hat{G}^\dagger=\hat{G}$ to go from the laboratory frame to a rotating frame. We choose 
\begin{align}
 \Hat{G}
 =&\sum_{m=1}^{N} \Big[\omega_1+\sum_{n
 < m}(-1)^{n+1}\omega^{s}_{n}\Big]\hat{\sigma}_{m,m},
\end{align}
where $\omega^{s}_{n}=\omega_{s}+\delta\omega_s$  $(\omega^{s}_{n}=\omega_{s})$ for odd (even) nonzero integer values of $n$. Applying the above transformation to the system's Hamiltonian $\hat{H}(t)$, we find the time-independent Hamiltonian $\hat{H}_{R}$ in the rotating frame, where [2]
\begin{align}
    \hat{H}_{R}= i \partial_{t} \hat{\mathcal{U}}(t)\hat{\mathcal{U}}^\dagger(t)+\hat{\mathcal{U}}(t)\hat{H}(t)\hat{\mathcal{U}}^\dagger(t)=-\hat{G}+\hat{\mathcal{U}}(t)\hat{H}(t)\hat{\mathcal{U}}^\dagger(t).
\end{align}
We find after some simplification:
\begin{align} 
\hat{H}_{R}
=\sum_{m= 1}^{N}\Big[\sum_{n<m}(-1)^{n+1}\Delta_{n}\Big]\hat{\sigma}_{m,m}+\sum_{m>1}^{N}\Omega\Big[\hat{\sigma}_{m,m-1}+\hat{\sigma}_{m-1,m}\Big],
\end{align}
where 
the detuning of the Raman fields from an optical transition is defined by $\Delta_n= (-1)^{n}\omega_{n,n+1}-\omega^{s}_{n}$.
 We can further relate the Heisenberg operators in the two frames using the relation: $\hat{U}_{R}(t)=\hat{\mathcal{U}}(t)\hat{U}(t)$. Here, $\hat{U}(t)$ and $\hat{U}_{R}(t)$ are the time evolution operators in the laboratory and the rotating frames, respectively. Therefore, we write for the  operators in the laboratory frame as
\begin{align}
    \hat{\sigma}_{m,m'}(t)=\hat{U}^\dagger(t) \hat{\sigma}_{m,m'}\hat{U}(t)=\hat{U}^\dagger_{R}(t) \hat{\mathcal{U}}(t)\hat{\sigma}_{m,m'}\hat{\mathcal{U}}^\dagger(t)\hat{U}_{R}(t).
\end{align}
Now,
\begin{gather}
    \hat{\mathcal{U}}(t)\hat{\sigma}_{m,m'}\hat{\mathcal{U}}^\dagger(t)=e^{-i\theta_{m,m'}t}\hat{\sigma}_{m,m'},~\text{where}~~\theta_{m,m'}=\sum_{n
 < m}(-1)^{n}\omega^{s}_{n}-\sum_{n
 < m'}(-1)^{n}\omega^{s}_{n}
   ~~~\Rightarrow~ \hat{\sigma}_{m,m'}(t)=e^{-i\theta_{m,m'}t}\hat{\sigma}^{R}_{m,m'}(t).\label{LabToRot}
\end{gather}
We denote the Heisenberg operators in the rotating frame by $\hat{\sigma}^{R}_{m, m'}(t)=\hat{U}^\dagger_{R}(t) \hat{\sigma}_{m, m'}\hat{U}_{R}(t)$.
In terms of these Heisenberg operators,  the Hamiltonian in the rotating frame can be written as follows:
\begin{align} 
\hat{H}_{R}(t)
=\sum_{m= 1}^{N}\Big[\sum_{n<m}(-1)^{n+1}\Delta_{n}\Big]\hat{\sigma}^{R}_{m,m}(t)+\sum_{m>1}^{N}\Omega\Big[\hat{\sigma}^{R}_{m,m-1}(t)+\hat{\sigma}^{R}_{m-1,m}(t)\Big].
\end{align}
We next discuss a microscopic system-bath model in the rotating frame.

\section{Microscopic system-bath model}
Following the manuscript, we now consider a microscopic system–bath model to incorporate the relaxation parameters $\gamma,\gamma\,'$ into our system. In the following, we reserve the symbols $p,q$ for even integers and $l,j$ for odd integers, respectively.
We consider that the bosonic baths are connected to all possible optical (electric-dipole) transitions between the ground and excited Zeeman states. These  bosonic modes of momentum $k$ with the creation and annihilation operators $\hat{a}^{\dagger}_{p,j}(k), \hat{a}_{p,j}(k)$, are coupled to the transition between an excited state $|p\rangle $ and a ground state $ |j\rangle$ with amplitude $g_{p,j}$. 
In addition, we consider another type of bosonic baths, with annihilation and creation operators $\hat{d}_{j,j'}(k)$ and $\hat{d}^\dagger_{j,j'}(k)$, which are coupled to transitions between two states $|j\rangle$ and $|j'\rangle$ in the ground manifold with amplitude $g'_{j,j'}$. 
The complete Hamiltonian comprising the system and the baths in the rotating frame is given by [3]
\begin{align}
   \hat{H}_{tot}(t)=&\sum_{p,j}\int dk ~\omega_k \hat{a}_{p,j}^\dagger(k,t)\hat{a}_{p,j}(k,t)+\sum_{j,j'}\int dk ~\omega_k' \hat{d}_{j,j'}^\dagger(k,t)\hat{d}_{j,j'}(k,t)+\hat{H}_R(t) +\sum_{p,j} g_{p,j}\int dk \Big[\hat{a}^\dagger_{p,j}(k,t)\hat{\sigma}^{R}_{j,p}(t)
   \nonumber\\
   &+ \hat{\sigma}^{R}_{p,j}(t) \hat{a}_{p,j}(k,t) \Big]+\sum_{j,j'}g'_{j,j' }\int dk~ \Big[\hat{d}_{j,j'}^\dagger(k,t) \hat{\sigma}^{R}_{j',j}(t)+\hat{\sigma}^{R}_{j,j'}(t)\hat{d}_{j,j'}(k,t)\Big],
\end{align}
 where $\omega_k,\omega_k'$ are the energy dispersion of the bosonic baths. We assume that the baths are connected to the system at time $t_o$. Using the Heisenberg evolution,  we get the following equations of motion for the bath operators : 
\begin{align}
    &\frac{\partial \hat{a}_{p,j}(k,t)}{\partial t}=-i\omega_k \hat{a}_{p,j}(k,t)-ig_{p,j}\hat{\sigma}^{R}_{j,p}(t)~\Rightarrow~ \hat{a}_{p,j}(k,t)=\hat{a}_{p,j}(k,t_o)e^{i\omega_k (t_o-t)}-ig_{p,j}\int_{t_o}^{t}e^{i\omega_k (\tau -t)}  \hat{\sigma}^{R}_{j,p}(\tau)d\tau,\\
    &\frac{\partial \hat{d}_{j,j'}(k,t)}{\partial t}=-i\omega_k' \hat{d}_{j,j'}(k,t)-ig'_{j,j'} \hat{\sigma}^{R}_{j',j}(t) ~\Rightarrow ~  \hat{d}_{j,j'}(k,t)=\hat{d}_{j,j'}(k,t_o)e^{i\omega_k' (t_o-t)}-ig'_{j,j'}\int_{t_o}^{t}e^{i\omega_k' (\tau -t)}\hat{\sigma}^{R}_{j',j}(\tau) d\tau.
\end{align}
 We further consider these baths to have linear energy-momentum dispersion with an inifinite bandwidth, i.e.,  $\omega_k=v_g k$, and $\omega'_k=v_g' k$ where $k\in (-\infty,\infty)$. This assumption helps us to simplify the time-integration in the system operator equations. We use
\begin{align}
    &\int_{t_o \to -\infty}^{t}\int dk ~e^{i\omega_k (\tau -t)} f(\tau)d\tau = \int_{t_o \to -\infty}^{t}\int dk ~e^{iv_g k (\tau -t)} f(\tau)d\tau=\frac{2\pi}{v_g} \int_{t_o \to \infty}^{t}\delta(\tau-t) f(\tau)d\tau=\frac{\pi}{v_g} f(t),
\end{align}
and similarly for the other integrals. We substitute the above solutions into the equations of motion for the atomic operators and obtain the quantum Langevin equations for the system operators as follows:
\begin{align}
   &\frac{\partial \hat{\sigma}^{R}_{q,l}(t)}{\partial t}
 =i[\hat{H}_R(t),\hat{\sigma}^{R}_{q,l}]+i\sum_{p} \sum_{j} g_{p,j}  \hat{\eta}^\dagger_{p,j}(t)\Big[\hat{\sigma}^{R}_{j,l}\delta_{p,q}-\hat{\sigma}^{R}_{q,p}\delta_{l,j}\Big]-i \sum_{j,j'}g'_{j,j'}\Big(\hat{\eta}_{j,j'}^\dagger(t)\hat{\sigma}^{R}_{q,j}\delta_{l,j'}+\hat{\sigma}^{R}_{q,j'} \hat{\eta}_{j,j'}(t)\delta_{l,j}\Big)\nonumber\\
  &~~~~~~~~~~~~~~~~~~~-\bigg(\sum_{j'}\frac{\pi g'^2_{l,j'}}{v_g'}+\sum_{j} \frac{\pi g_{q,j}^2}{v_g}\bigg)\hat{\sigma}^{R}_{q,l}~~~\text{for}~~q=\text{even, and  } l=\text{odd},\label{L1}
  \end{align}
  \begin{align}
 &\frac{\partial \hat{\sigma}^{R}_{l,l}(t)}{\partial t}=i[\hat{H}_R(t),\hat{\sigma}^{R}_{l,l}]+i\sum_{j,j'} g'_{j,j'}\Big[\hat{\eta}_{j,j'}^\dagger(t) \big(\hat{\sigma}^{R}_{j',l}\delta_{j,l}-\hat{\sigma}^{R}_{l,j}\delta_{l,j'}\big)+\big(\hat{\sigma}^{R}_{j,l}\delta_{j',l}-\hat{\sigma}^{R}_{l,j'}\delta_{l,j}\big)\hat{\eta}_{j,j'}(t)\Big]+i\sum_{p}\Big[ g_{p,l} \hat{\sigma}^{R}_{p,l} \hat{\eta}_{p,l}(t)    \nonumber\\
  &~~~~~~~~~~~~~~~~~~~~-g_{p,l}\hat{\eta}^\dagger_{p,l}(t)\hat{\sigma}^{R}_{l,p}\Big]+\sum_{p} \frac{2\pi g_{p,l}^2}{v_g}\hat{\sigma}^{R}_{p,p}-\sum_{j'} \frac{2\pi g'^2_{l,j'} }{v_g'}\hat{\sigma}^{R}_{l,l}+\sum_{j} \frac{2\pi g'^2_{j,l}}{v_g'}\hat{\sigma}^{R}_{j,j}~~~\text{for}~~l=\text{odd},\label{L2}\\
  &\frac{\partial \hat{\sigma}^{R}_{l,l'}(t)}{\partial t}=i[\hat{H}_R(t),\hat{\sigma}^{R}_{l,l'}]+i\sum_{j,j'} g'_{j,j'}\Big[\hat{\eta}_{j,j'}^\dagger(t) \big(\hat{\sigma}^{R}_{j',l'}\delta_{j,l}-\hat{\sigma}^{R}_{l,j}\delta_{l',j'}\big)+\big(\hat{\sigma}^{R}_{j,l'}\delta_{j',l}-\hat{\sigma}^{R}_{l,j'}\delta_{l',j}\big)\hat{\eta}_{j,j'}(t)\Big]+i\sum_{p}\Big[ g_{p,l} \hat{\sigma}^{R}_{p,l'} \hat{\eta}_{p,l}(t)    \nonumber\\
  &~~~~~~~~~~~~~~~~~~~~-g_{p,l'}\hat{\eta}^\dagger_{p,l'}(t)\hat{\sigma}^{R}_{l,p}\Big]-\sum_{j'} \frac{\pi }{v_g'}(g'^2_{l,j'}+g'^2_{l',j'})\hat{\sigma}^{R}_{l,l'}~~~\text{for}~~l,l'~\text{are odd, and }~l\neq l',\label{L3}\\
  &\frac{\partial \hat{\sigma}^{R}_{q,q'}(t)}{\partial t}=i[\hat{H}_R(t),\hat{\sigma}^{R}_{q,q'}]-i\sum_{p}\sum_{j} g_{p,j} \Big[\delta_{q',p}\hat{\sigma}^{R}_{q,j} \hat{\eta}_{p,j}(t)- \delta_{p,q} \hat{\eta}^\dagger_{p,j}(t)\hat{\sigma}^{R}_{j,q'}\Big]-\sum_{j} \frac{\pi }{v_g}\Big[g_{q',j}^2  +g_{q,j}^2 \Big]\hat{\sigma}^{R}_{q,q'}~~\text{for}~~q,q'=\text{even}.\label{L4}
\end{align}
We have introduced noises from the baths in the above quantum Langevin equations, and the noises are
\begin{align}
    \hat{\eta}_{j,j'}(t)=\int dk~ e^{i\omega'_k (t_o -t)}\hat{d}_{j,j'}(k,t_o),~~~\text{and}~ \hat{\eta}_{p,j}(t)=\int dk~ e^{i\omega_k (t_o -t)}\hat{a}_{p,j}(k,t_o).
\end{align}
We now proceed with one more assumption: we drop the noises from Eqs.~\ref{L1}-\ref{L4} assuming that their contributions are negligible for the strongly driven system.  The resulting equations of motion for the system operators in the rotating frame read as:
\begin{align}
      \frac{\partial \hat{\sigma}^{R}_{m,m'}(t)}{\partial t}  =&i\bigg[\sum_{n<m'}(-1)^{n}\Delta_{n}-\sum_{n<m}(-1)^{n}\Delta_{n}\bigg]\hat{\sigma}^{R}_{m,m'}(t)+i\Omega\bigg[\hat{\sigma}^{R}_{m+1,m'}(t)-\hat{\sigma}^{R}_{m,m'-1}(t)+\hat{\sigma}^{R}_{m-1,m'}(t)-\hat{\sigma}^{R}_{m,m'+1}(t)\bigg]\nonumber\\
     &-\sum_{n}\bigg[\frac{\gamma\,'_{m,n}+\gamma\,'_{m',n}}{2}\hat{\sigma}^{R}_{m,m'}(t)-\gamma\,'_{n,m}\delta_{m,m'}\hat{\sigma}^{R}_{n,n}(t) \bigg]-\sum_{n}\bigg[\frac{\gamma_{m,n}+\gamma_{m',n}}{2}\hat{\sigma}^{R}_{m,m'}(t)-\gamma_{n,m}\delta_{m,m'}\hat{\sigma}^{R}_{n,n}(t) \bigg],\label{QLE_rot}
\end{align}
where $\gamma\,'_{m,n}=2\pi g\,'^2_{m,n}/v_g'$ is nonzero for odd $m,n$, and $\gamma_{n,m}=2\pi g^2_{n,m}/ v_g$ is nonzero only when $n$ is even and $m$ is odd. 

We now perform the expectation $\langle \dots \rangle$ of the operators in the above Eq.~\ref{QLE_rot} in the initial state of the system and baths. We define $\rho^{R}_{m',m}(t)=\langle \hat{\sigma}^{R}_{m,m'}(t)\rangle$ and get the following equation from Eq.~\ref{QLE_rot}:
    \begin{align}
\frac{\partial {\rho}^{R}_{m',m}(t)}{\partial t}  =&i\bigg[\sum_{n<m'}(-1)^{n}\Delta_{n}-\sum_{n<m}(-1)^{n}\Delta_{n}\bigg]{\rho}^{R}_{m',m}(t)+i\Omega\bigg[{\rho}^{R}_{m',m+1}(t)-{\rho}^{R}_{m'-1,m}(t)+{\rho}^{R}_{m',m-1}(t)-{\rho}^{R}_{m'+1,m}(t)\bigg]\nonumber\\
     &-\sum_{n}\bigg[\frac{\gamma\,'_{m,n}+\gamma\,'_{m',n}}{2}{\rho}^{R}_{m',m}(t)-\gamma\,'_{n,m}\delta_{m,m'}{\rho}^{R}_{n,n}(t) \bigg] -\sum_{n}\bigg[\frac{\gamma_{m,n}+\gamma_{m',n}}{2}{\rho}^{R}_{m',m}(t)-\gamma_{n,m}\delta_{m,m'}{\rho}^{R}_{n,n}(t)\bigg].\label{master_rot}
\end{align}
The above equation has a unique steady state at long times $(t\gg \gamma\,'^{-1})$, and we denote  its value by $\rho^{R}_{m',m}$. We refer to $\rho^{R}_{m',m}$ as the population and the coherence for $m=m'$ and $m \neq m'$, respectively.

The power spectrum, $P_{l}^{(n)}(\omega)$, defined in the manuscript, is simplified using the rotating frame operators as follows:
\begin{align}
   P_{l}^{(n)}(\omega)=\frac{1}{2\pi} \int_{-\infty}^{\infty} d\tau~ e^{i\omega \tau} ~ \langle \hat{\sigma}^{\dagger}_{l,l+2n}(t) \hat{\sigma}_{l,l+2n} (t+\tau) \rangle=&\frac{1}{2\pi} \int_{-\infty}^{\infty} d\tau~ e^{i(\omega-n\delta\omega_s) \tau} ~ \langle \hat{\sigma}^{R\,\dagger}_{l,l+2n}(t) \hat{\sigma}^{R}_{l,l+2n} (t+\tau) \rangle.
\end{align}
We use $\theta_{l,l+2n}=n\delta\omega_s$. As we discuss in the paper, $P_{l}^{(n)}(\omega)$ consists of coherent and incoherent parts. For the coherent part, we get the following at steady state [3,4]:
 \begin{align}
 \lim_{t \to \infty}\langle \hat{\sigma}^{R\,\dagger}_{l,l+2n}(t) \hat{\sigma}^{R}_{l,l+2n} (t+\tau) \rangle=\lim_{t \to \infty}\langle \hat{\sigma}^{R\,\dagger}_{l,l+2n}(t) \rangle \lim_{t \to \infty}\langle\hat{\sigma}^{R}_{l,l+2n} (t+\tau) \rangle=|\rho^{R}_{l+2n,l}|^2.
 \end{align}

\section{Effective non-Hermitian description}
In the manuscript, we use an effective non-Hermitian description of the system involving the degrees of freedom of the Zeeman states in the ground hyperfine manifold. Here, we provide the detailed derivation of Eq.~9 that is used in the manuscript. 
We consider the population of the ground Zeeman states decays into its nearest neighbor states in the ground manifold at a rate of $\gamma\,'$. 
We further assume that the population of each excited state $|m\rangle$ decays to its adjacent ground states $|m\pm 1\rangle$ at a rate of $\gamma$. Therefore, $\gamma_{m,n}=\gamma~  \delta_{n,m\pm 1}$ for even $m$, or else $\gamma_{m,n}=0$ in Eq.~\ref{QLE_rot}.
With these assumptions, the coherence in the ground states is governed by the equation
\begin{align}
     \frac{\partial }{\partial t} \hat{\sigma}^{R}_{l,l'}(t)=&\bigg[i\sum_{n<l'}(-1)^{n}\Delta_{n}-i\sum_{n<l}(-1)^{n}\Delta_{n}-\frac{4-\delta_{l,1}-\delta_{l,N}-\delta_{l',1}-\delta_{l',N}}{2}\gamma\,'\bigg]\hat{\sigma}^{R}_{l,l'}(t)\nonumber\\
     &+i\Omega\Big[\hat{\sigma}^{R}_{l+1,l'}(t)-\hat{\sigma}^{R}_{l,l'-1}(t)+\hat{\sigma}^{R}_{l-1,l'}(t)-\hat{\sigma}^{R}_{l,l'+1}(t)\Big]~~\text{for}~l\neq l',\label{S21}
\end{align}
and their population dynamics is determined using 
\begin{align}
     \frac{\partial }{\partial t} \hat{\sigma}^{R}_{l,l}(t)=&i\Omega\Big[\hat{\sigma}^{R}_{l+1,l}(t)-\hat{\sigma}^{R}_{l,l-1}(t)+\hat{\sigma}^{R}_{l-1,l}(t)-\hat{\sigma}^{R}_{l,l+1}(t)\Big]+\gamma(\hat{\sigma}^{R}_{l-1,l-1}+\hat{\sigma}^{R}_{l+1,l+1})\nonumber\\
     &-(2-\delta_{l,1}-\delta_{l,N})\gamma\,'\hat{\sigma}^{R}_{l,l}+ \gamma\,'(\hat{\sigma}^{R}_{l-2,l-2}+\hat{\sigma}^{R}_{l+2,l+2}).\label{S22}
\end{align}
We integrate out the fast relaxing dynamics $(\gamma \gg \gamma\,')$  of the excited states assuming [5]:
\begin{align}
    \frac{\partial }{\partial t}  \hat{\sigma}^{R}_{l-1,l'}(t) \approx 0, ~~\text{and} ~~\frac{\partial }{\partial t}  \hat{\sigma}^{R}_{l-1,l'-1}(t) \approx 0,\label{Eq_Dark}
\end{align}
 at times $t \gg \gamma^{-1}$.  From the above equations, we get:
\begin{gather}
    \hat{\sigma}^{R}_{l-1,l'}(t) \approx-\frac{\Omega\big[\hat{\sigma}^{R}_{l,l'}(t)+\hat{\sigma}^{R}_{l-2,l'}(t)\big]}{\sum_{n<l'}(-1)^{n}\Delta_{n}-\sum_{n<l-1}(-1)^{n}\Delta_{n}+i\gamma} \approx -\frac{\Omega}{i\gamma}\big[\hat{\sigma}^{R}_{l,l'}(t)+\hat{\sigma}^{R}_{l-2,l'}(t)\big], \label{S19}\\
   \text{and}~~~  \hat{\sigma}^{R}_{l-1,l-1} (t)=\frac{i\Omega}{2\gamma}\big[\hat{\sigma}^{R}_{l,l-1}(t)-\hat{\sigma}^{R}_{l-1,l-2}(t)+\hat{\sigma}^{R}_{l-2,l-1}(t)-\hat{\sigma}^{R}_{l-1,l}(t)\big].\label{S20}
\end{gather}
We take $\gamma > \Omega>\Delta_n$ for all $n$. On the right-hand side of Eq.~\ref{S19}, we drop the operators governing the coherence and populations in the excited manifold as they contribute less than the other existing terms when $\Omega/\gamma < 1$. 
We substitute the approximate solutions obtained from Eqs.~(\ref{S19}-\ref{S20}) in Eq.~\ref{S21} and Eq.~\ref{S22}, and write the effective equations of motion in the rotating frame as follows:
\begin{align}
    &\frac{\partial }{\partial t}\hat{\sigma}^{R}_{l,l'}(t)=i\big[\hat{H}_{R,\text{eff}}(t)\hat{\sigma}^{R}_{l,l'}(t)-\hat{\sigma}^{R}_{l,l'}(t)\hat{H}^\dagger_{R,\text{eff}}(t)\big](1-\delta_{l,l'})- \tilde{\gamma}^{\,\text{eff}}_{l,l'}(1-\delta_{l,l'})\hat{\sigma}^{R}_{l,l'}(t)-\sum_{j}\bigg[\frac{\gamma\,'^{\text{eff}}_{l, j}+\gamma\,'^{\text{eff}}_{l',j}}{2}\hat{\sigma}^{R}_{l,l'}(t)-\gamma\,'^{\text{eff}}_{j,l}\delta_{l,l'}\hat{\sigma}^{R}_{j,j}(t)\bigg].\label{Effrot_Eq}
\end{align}
Here $l,l'$ and $j$ are odd indices, and 
\begin{align} 
\hat{H}_{R,\text{eff}}(t)=\sum_{l}\Big[\sum_{n<l}(-1)^{n+1}\Delta_{n}\Big]\hat{\sigma}^{R}_{l,l}(t)+iJ_o\sum_{l\geq 3}\Big[\hat{\sigma}^{R}_{l,l-2}(t)+\hat{\sigma}^{R}_{l-2,l}(t)\Big].
\end{align}
Here, we obtain the effective decay rate for the population in the ground manifold as $\gamma\,'^{\text{eff}}_{l,l'}=J_o+\gamma\,'$ if $|l-l'|=2 $,  or else $\gamma\,'^{\text{eff}}_{l,l'}=0$ for other values of $l,l'$. Eliminating the excited states also leads to an additional decay term in the coherence dynamics between the ground Zeeman states, which is denoted by $\tilde{\gamma}^{\text{eff}}_{l,l'}$ in  Eq.~\ref{Effrot_Eq} and reads as $\tilde{\gamma}^{\text{eff}}_{l,l'}=(4-\delta_{l,1}-\delta_{l,N}-\delta_{l',1}-\delta_{l',N})J_o/2$.
Transformation of the Eq.~\ref {Effrot_Eq} back to the laboratory frame is carried out using: $\hat{\sigma}_{l,l'}(t)=e^{-i\theta_{l,l'}t}\hat{\sigma}^{R}_{l,l'}(t)$ and this gives Eq.~9 of the manuscript.
We fix $\Delta_1=\Delta$ and $\Delta_2=0$, and solve for the steady-state values of $\rho^{R}_{l',l}$ obtained using the effective non-Hermitian description described in Eq.~\ref{Effrot_Eq}.
In the following, we obtain a few analytical expressions for $\rho^{R}_{l',l}$ for $N=3,5,$ and $7$. 
\subsection{3-level cascaded $\Lambda$ system (3-LC$\Lambda$S)}
We first consider $N=3$, where the Eq.~\ref{Effrot_Eq} involves two Zeeman states in the ground manifold. The coherence between these two ground states in the steady state is given by 
\begin{align}
    \rho^{R}_{1,3}= -\frac{J_{o}}{(\gamma\,'+2 J_{o})-i\Delta}.
\end{align}
The steady-state  populations are:  $ \rho^{R}_{l,l}= 1/2$ for $l=1$, $3$. Here, the power spectrum of coherence shows a single peak at $\omega=\delta\omega_s$. 
\subsection{5-level cascaded $\Lambda$ system (5-LC$\Lambda$S)}
Next, we consider $N=5$.  The coherence between the ground Zeeman states take complicated relations as follows: \begin{align}
    &\rho^{R}_{1,3}=-\frac{4 }{3}\frac{J_o(2J_o+\gamma\,'-2i\Delta)}{8 J_o^2+12 J_o \gamma\,'+3\gamma\,'^2-8i(2J_o+\gamma\,')\Delta-4\Delta^2},\nonumber\\
    &\rho^{R}_{1,5}=\frac{8}{3}\frac{J_o^{2}}{8 J_o^2+12 J_o \gamma\,'+3\gamma\,'^2-8i(2J_o+\gamma\,')\Delta-4\Delta^2},
\end{align}
and $\rho^{R}_{3,5}=\rho^{R}_{1,3}$. The steady-state  populations give : $ \rho^{R}_{l,l}= 1/3$. In the limit of small $J_o (\ll \gamma\,')$, we observe the following variations of the nearest-neighbor coherence $\rho^{R}_{1,3} \simeq -4J_o/(9\gamma\,')$ and the next nearest-neighbor coherence $\rho^{R}_{1,5} \simeq 8J_o^2/(9\gamma\,'^2)$, where $\Delta=0$. In the opposite limit, when $J_o \gg \gamma\,',\Delta$, we get the saturation values : $\rho^{R}_{1,3} \approx -1/3$ and $\rho^{R}_{1,5} \approx 1/3$. In the power spectrum, we find two contributions  $P^{(1)}_{1}(\omega), P^{(1)}_{3}(\omega)$ for the fundamental peak at $\omega=\delta\omega_s$ and only one contribution $P^{(2)}_{1}(\omega)$ for the first harmonics at $\omega=2\omega_s$. The corresponding height ratio of the two coherent peaks is given in the manuscript.
\subsection{7-level cascaded $\Lambda$ system (7-LC$\Lambda$S)}
To inspect next-to-next neighbor coherence between the Zeeman states in the ground manifold, we take $N=7$. 
At steady state, we get $ \rho^{R}_{l,l}= 1/4$. Using the leading-order nonzero contributions in the coherence for $J_o \ll \gamma\,',\Delta$, we obtain the following. 
\begin{gather}
    \rho^{R}_{1,3}\simeq -\frac{J_o}{3\gamma\,'-2i\Delta},~~
    \rho^{R}_{1,5}\simeq\frac{J^{2}_o (7\gamma\,'-4i\Delta)}{18\gamma\,'^3-45 i\gamma\,'^2\Delta-34 \gamma\,' \Delta^2+8i\Delta^3},\\
    \rho^{R}_{1,7}\simeq-\frac{2J^{3}_o (7\gamma\,'-4i\Delta)}{18\gamma\,'^4-99 i\gamma\,'^3 \Delta-169 \gamma\,'^2\Delta^2+110i \gamma\,' \Delta^3+24 \Delta^4},
\end{gather}
with $\rho^{R}_{1,3}=\rho^{R}_{3,5}=\rho^{R}_{5,7}$, and $\rho^{R}_{1,5}= \rho^{R}_{3,7}$. Additionally, when $\Delta=0$, the above expressions give:
$\rho^{R}_{1,3}\simeq -J_o/(3\gamma\,')$, $\rho^R_{1,5}\simeq 7J_o^2/(18 \gamma\,'^2)$, and $\rho^R_{1,7}\simeq- 7J_o^3/(9 \gamma\,'^3)$.
 In this system, we get three coherent peaks in the power spectrum that characterize the coherence between the four Zeeman states in the ground manifold. 
 With the above solutions, we find the height ratios at smaller values of $J_o$, which are provided in the manuscript.
In the opposite limit, $J_o \gg \Delta, \gamma\,'$, the coherence saturate to the following values $\rho^{R}_{1,3}=\rho^{R}_{1,7}=\rho^{R}_{3,5}=\rho^{R}_{5,7}\approx-1/4$, and $\rho^{R}_{1,5}= \rho^{R}_{3,7}\approx 1/4$.

\section{Details of lasers and opto-mechanical parts used in the experiment}
The experiment is performed in a cylindrical glass cell of length $75$ mm and diameter $19$ mm. The frequencies of the lasers are monitored using a wavelength meter (WSU-2, HighFinesse GmbH) with an absolute accuracy of $\pm1$ MHz. All laser sources are external cavity diode lasers (ECDL, Toptica Photonics). A polarizing beam splitter (PBS122) and a low-noise 100 MHz bandwidth balanced photodetector (PDB230A) are used for the polarimetric detection. The spectral measurements are carried out with a spectrum analyzer (FPC1500). Temperature control of the cell is achieved using polyimide film flexible heaters (OMEGA KHLVA-105/10-P) in combination with a PID controller (OMRON E5CWL). A low-noise current source drives the magnetic field coils (LD3000, Thorlabs). Acousto-optic modulators (ISOMET 1206C and AA OPTICS MT110-A1-IR) are incorporated into the setup and controlled by our PXIe system.
\bigskip
\hrule 
\bigskip

\noindent[1] E. N. Economou, \href{https://link.springer.com/book/10.1007/3-540-28841-4}{Green’s Functions in Quantum Physics } (Springer, 2006).

\noindent[2] D. A. Steck, \href{https://atomoptics.uoregon.edu/~dsteck/teaching/quantum-optics/}{Quantum and atom optics} (Springer, revision 0.8.3, 25 May 2012).

\noindent[3] D. Roy, Critical features of nonlinear optical isolators for improved nonreciprocity, \href{https://journals.aps.org/pra/abstract/10.1103/PhysRevA.96.033838}{Phys. Rev. A 96, 033838 (2017).}

\noindent[4] K. Koshino and Y. Nakamura, Control of the radiative level shift and linewidth of a superconducting artificial atom through a variable
boundary condition, \href{https://iopscience.iop.org/article/10.1088/1367-2630/14/4/043005}{ New J. Phys. 14, 043005 (2012).}

\noindent[5] L. Zhang, F. Yang, K. Mølmer, and T. Pohl, Unidirectional quantum-optical elements for waveguide-QED with subwavelength Rydberg-
atom arrays in free space, \href{https://opg.optica.org/opticaq/fulltext.cfm?uri=opticaq-3-3-256&id=572991}{Opt. Quantum. 3, 256 (2025).}

\end{document}